\documentclass[letterpaper, 10 pt, conference]{ieeeconf}  

\IEEEoverridecommandlockouts                              
\usepackage{amsmath,amsfonts}
\usepackage{algorithmic}
\usepackage{array}
\usepackage[caption=false,font=normalsize,labelfont=sf,textfont=sf]{subfig}
\usepackage{textcomp}
\usepackage{xcolor}
\usepackage{stfloats}
\usepackage{url}
\usepackage{verbatim}
\usepackage{graphicx}
\def\BibTeX{{\rm B\kern-.05em{\sc i\kern-.025em b}\kern-.08em
    T\kern-.1667em\lower.7ex\hbox{E}\kern-.125emX}}
\usepackage{balance}                      

\title{\LARGE \bf
Excitation of Looped Bistable Bands for High-Speed Linear Actuation}

\author{Sareum Kim and Josie Hughes
\thanks{$^{1}$ All authors are with the CREATE Lab, Institute of Mechanical Engineering, Swiss Federal Institute of Technology in Lausanne (EPFL), 1015 Lausanne, Switzerland.
       (email: {\tt\small sareum.kim@epfl.ch, josie.hughes@epfl.ch})}
}

\begin{document}

\maketitle
\thispagestyle{empty}
\pagestyle{empty}

\begin{abstract}

Soft robotics increasingly relies on smart materials and innovative structures, with bistable tape springs emerging as a promising option. These structures exhibit intriguing dynamic behaviors, such as oscillation, due to their inherent bistability. This paper explores the high-speed linear amplification of motion achieved through the excitation of a looped bistable tape spring. When looped, the tape spring forms two distinct joints, facilitating smooth oscillation. Mounted on a linear guide and driven by a crank mechanism with varying frequency, the system converts input oscillations into amplified linear motion at resonance. This study highlights the potential of bistable tape springs high speed reciprocating linear motion. 

\end{abstract}

\section{INTRODUCTION}
Soft robotics, characterized by its flexibility and adaptability, demands actuators that incorporate smart materials and structures. Bistable structures of wide interest for soft robotic applications due to their energy efficiency, rapid actuation, and mechanical stability.  One such widely used structure is a tape spring, a bistable structure which features a curved beam design providing flexibility and high flexural rigidity, making it well-suited for applications requiring stiffness and load-bearing capabilities. Consequently, it has been used in various soft robotic mechanisms, such as extensible arm for manipulation \cite{osele2022lightweight} and robot anchoring for locomotion \cite{reachbot,EEWOC}, surgical needles \cite{steerableneedle}, and lightweight deployable structures for wings \cite{ladybeetle} and space applications \cite{flexiblehinge}.

Another characteristic of tape springs is their smooth rolling behavior, where localized folding acts as a rotational joint and straight segments serve as linkages. Previous research has exploited this principle to make variable length manipulators  \cite{closedlooptapespring},\cite{he2023thesis},\cite{sparks2024pressure}, and rolling hinge joints\cite{rollinghinge}.  This can be extended by forming the tap into a loop, where two localized folding joint are made at each end of the loop. The unique joint formation in a looped tape spring enables smooth linear motion parallel to its length and smooth transitions between stable states, creating potential for high-speed resonant oscillitory motion.  

In this paper, we investigate the oscillatory behavior of a looped bistable tape spring excited by a linear motion. Oscillation characteristics depend on the excitation frequency and length of the loop.  With adequate design and control parameters, the loop can have amplify linear displacement through a resonant behaviour. We demonstrate a maximum speed of the loop motion of 121.3 cm/s. This results highlights the potential of looped bistable band as efficient and functional structure for soft robot.

\section{Oscillation of Looped Bistable Band}

\subsection{Structure of Looped Bistable Band}

\begin{figure}[!b]
\vspace{-0.5cm}
\centering
\includegraphics[width=\linewidth, trim={1cm 0.3cm 0 0, clip}]{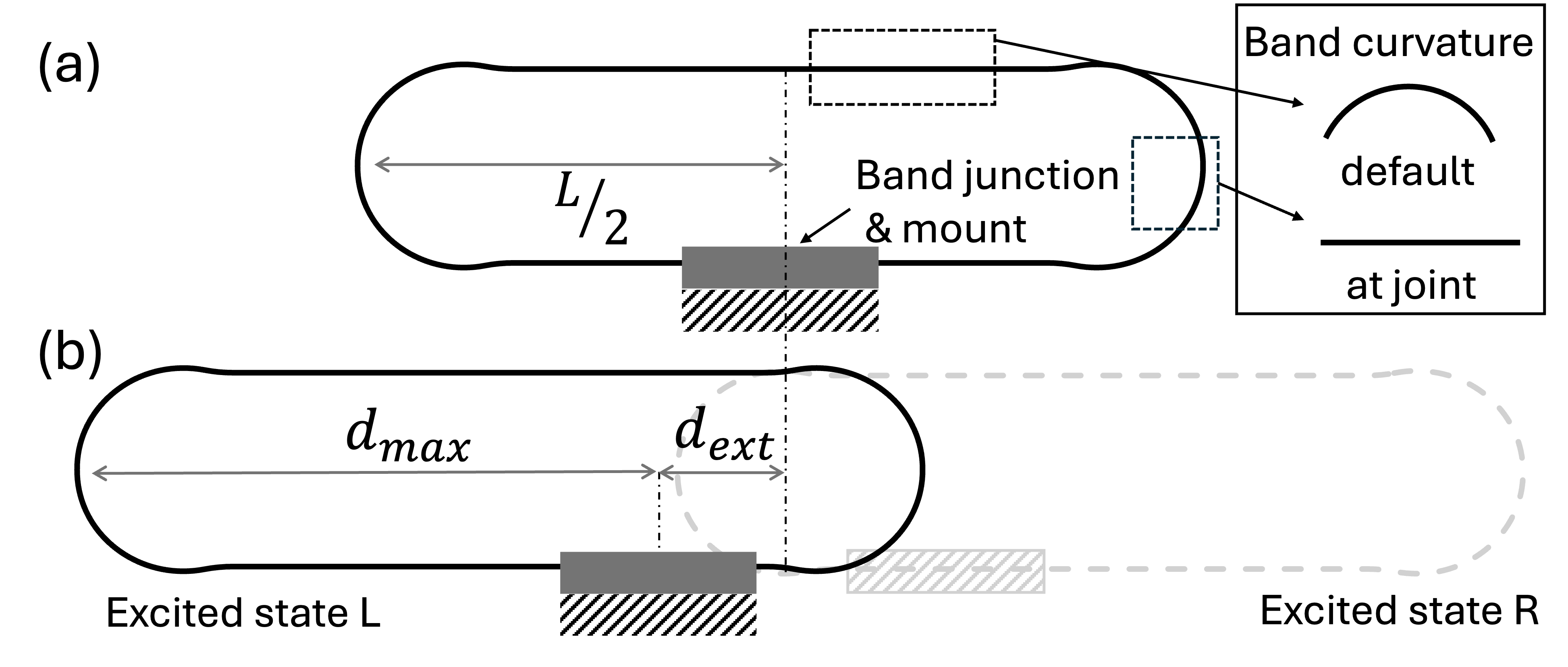}
\caption{Displacement amplification of looped bistable band (a) loop at stationary state (b) excited loop configuration after reaching excited state}
\label{fig:schematic}
\end{figure}

A looped bistable band made out of tape spring is shown in Fig. ~\ref{fig:schematic} (a). When the tape spring is made into a loop, there are two folding joints connected by flat tape, forming an oblong shape with characteristic length of $L$. The cross-section of the beam at the folding joint is flat, leading to a higher energy level, while the straight part remains a curved beam. Locating in the centre of the loop, the band junction is where two ends of the band are joined, and leads to a discontinuity of its characteristic. To induce oscillation in the loop, it is mounted on a linearly oscillating structure at the band junction. 

Under certain actuated conditions condition, the loop oscillates and reaches its maximum displacement as shown in Fig. ~\ref{fig:schematic} (b). Because of the joint cannot pass the band junction, the point of discontinuity, the band oscillates between two extreme states. The excitation distance in each direction (either excitation state L or R) is $d_{ext}$, and the maximum displacement of the loop is $d_{max}$. Then the amplified displacement ($d_{amp}=d_{max}-L/2$) is the difference between the maximum displacement and half of the original length of the loop. 

\vspace{-0.25cm}

\subsection{Excitation Setup}

\begin{figure}[!t]
\vspace{0.2cm}
\centering
\includegraphics[width=0.95\linewidth, trim={0.5cm 0.3cm 0 0, clip}]{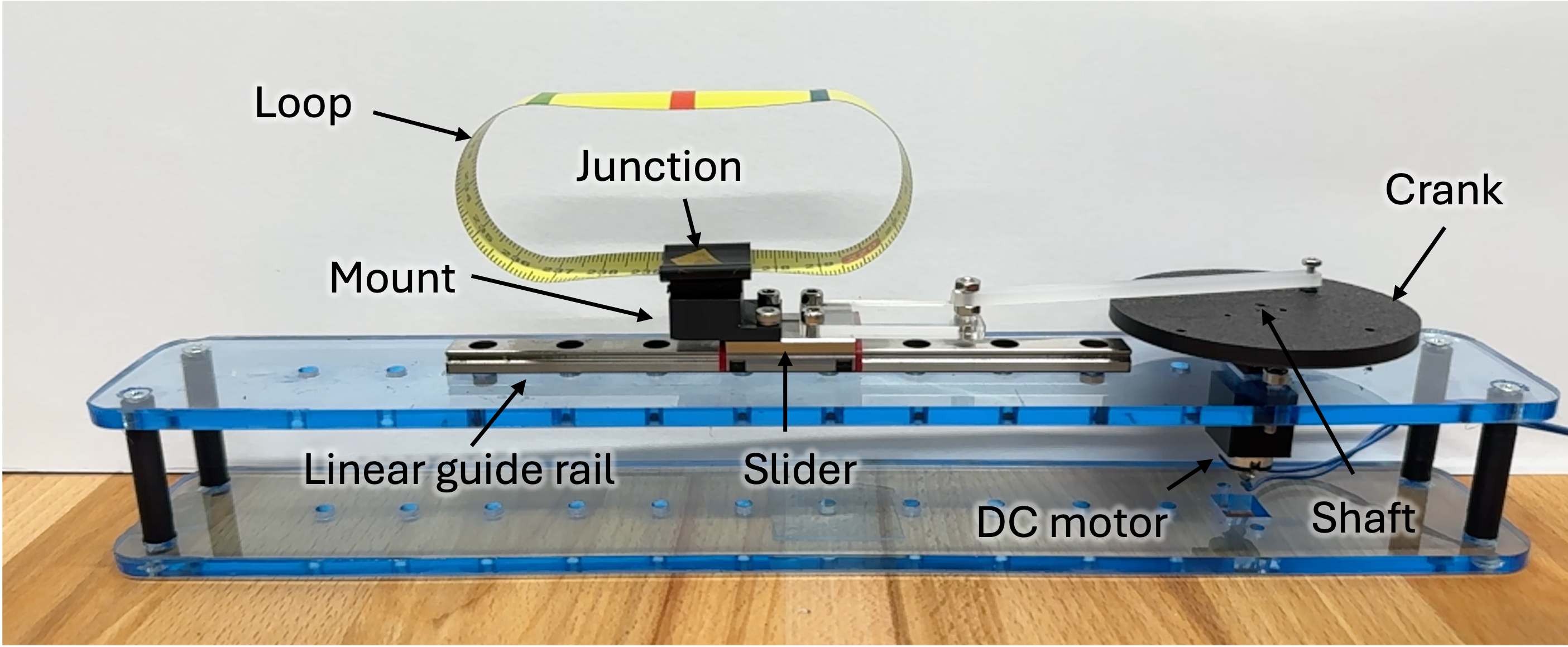}
\caption{Experimental setup for loop excitation induced by linear oscillation of crank mechanism}
\label{fig:experiment}
\end{figure}

Figure ~\ref{fig:experiment} shows the experimental setup for excitation. Three loop samples of different sizes (10 cm, 12.5 cm, and 15 cm) were tested. A slider mounted on the linear guide was driven by a DC motor (Pololu 75:1 Micro Metal Gearmotor), with a fixed excitation distance ($d_{ext}$=2 cm). By varying the PWM control with a DRV8838 motor driver, the excitation frequency was adjusted to seven different values between 2.3 Hz and 9.3 Hz. The maximum displacement was measured and averaged across three samples through analysis of recorded high-speed video.

\section{Result}
\subsection{Oscillation Profile} 
\begin{figure}[!b]
\vspace{-0.5cm}
\centering
\includegraphics[width=0.9\linewidth, trim={0.5cm 0.8cm 0.5cm 0cm, clip}]{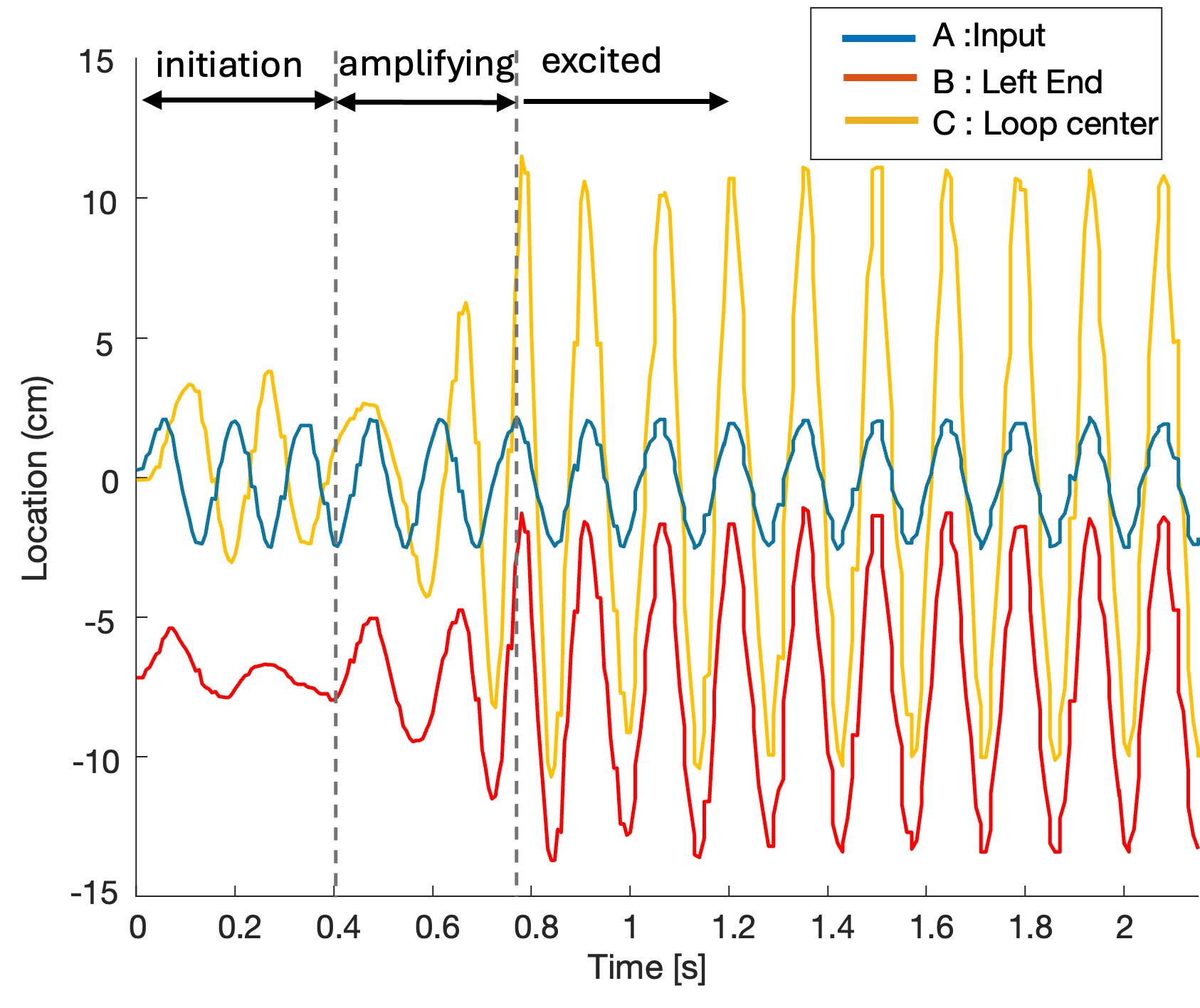}
\caption{Amplified distance and ratio according to excitation frequency for different loop sizes.}
\label{fig:oscillation}
\end{figure}

Fig. ~\ref{fig:oscillation} shows the oscillation profile of a loop (size $L$ = 12.5 cm, at 7.4 Hz) observed in three different location of the loop. Point 'A' represents the input tracked at the center of the slider, point 'B' denotes the far left position of the loop for measuring maximum displacement, and point 'C' corresponds the center of the loop.
The oscillation of the loop consists of three phases. First, when the mount begins to oscillate, the loop also starts oscillating with a small displacement in opposite phase to the mount; this is the initiation phase. Second, the oscillation aligns with the input oscillation phase and amplifies its displacement (amplification phase). Lastly, the loop reaches its maximum oscillation displacement and the motion stabilizes (excited phase).

\subsection{Frequency Effect}

\begin{figure}[!t]
\centering
\includegraphics[width=0.9\linewidth, trim={0.5cm 0.9cm 0.5cm 0cm, clip}]{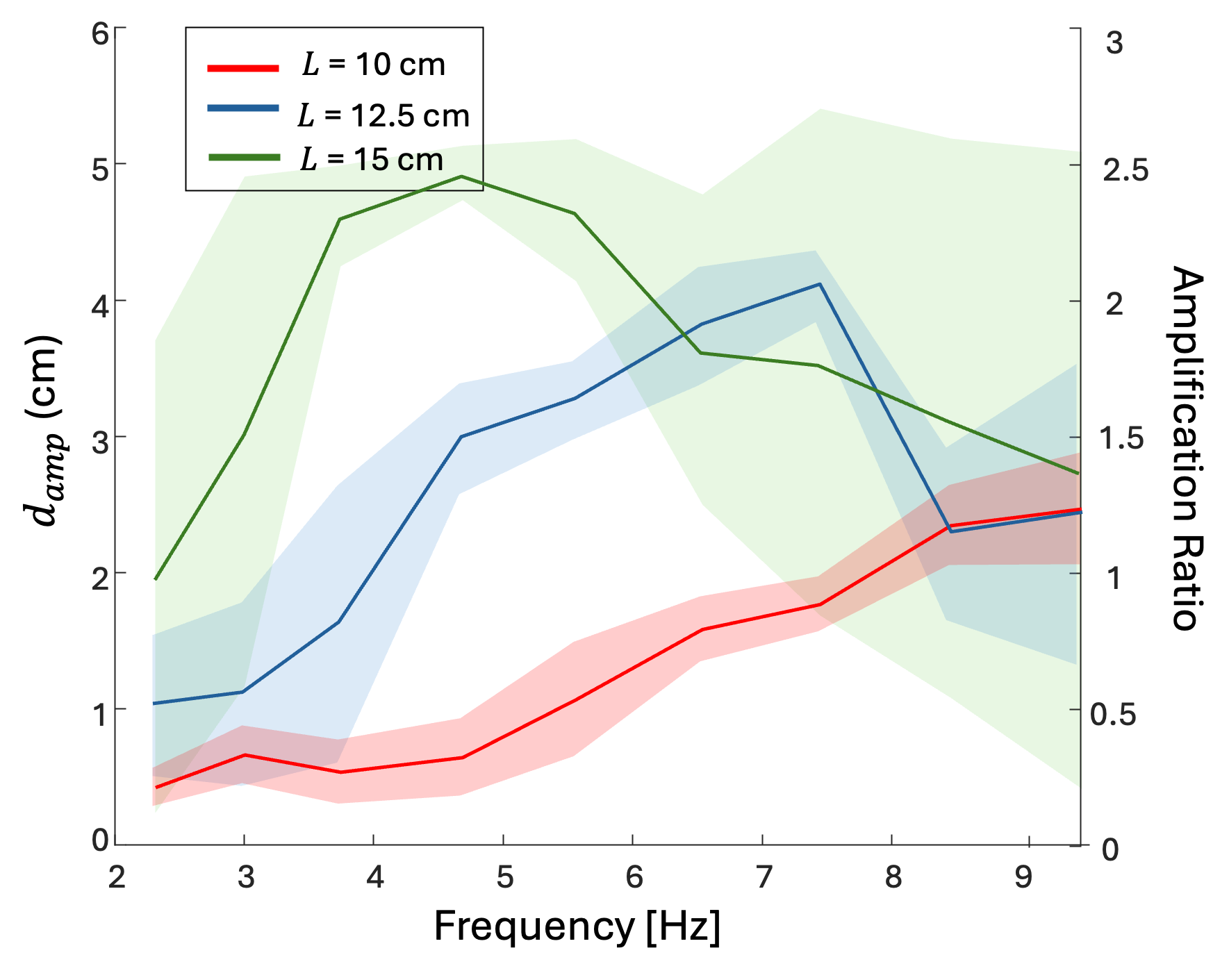}
\caption{Amplified distance and ratio according to excitation frequency for different loop sizes.}
\label{fig:excitation}
\end{figure}

As shown in Fig. ~\ref{fig:excitation}, the excitation frequency corresponds to maximum displacement decreases with increasing size of the loop.  Specifically, for the shortest loop length ($L$ = 10 cm), maximum displacement occurs at 9.4 Hz; for the middle length ($L$ = 12.5 cm), at 7.4 Hz; and for the longest length ($L$ = 15 cm), at 4.7 Hz. The maximum speed of loop is 121.3 cm/s for middle size loop ($L$ = 12.5 cm) at 7.4 Hz.

\section{CONCLUSIONS}

This study investigated the dynamic behavior of bistable tape springs in response to varying excitation frequencies and loop lengths. The experimental results demonstrated that longer loops exhibited lower excitation frequencies for maximum displacement, highlighting the relationship between characteristic length and oscillation behavior. These findings contribute to advancing our understanding of bistable tape springs and offer insights into optimizing their application for flexible and adaptive robotic systems.

\addtolength{\textheight}{-12cm}   









\end{document}